%% file: ex_article.tex
\documentclass[review,onefignum,onetabnum]{siamart171218}

\input{ex_shared}

\ifpdf
\hypersetup{
  pdftitle={General algorithm of assigning raster features to vector map at any resolution or scale},
  pdfauthor={N. Xu, M. Stevenson, K. Nice and S.Seneviratne}
}
\fi

\myexternaldocument{ex_supplement}

\usepackage{amsmath}
\usepackage{amsfonts}
\usepackage{amssymb}
\usepackage{mathtools}

\begin{document}
\nolinenumbers
\maketitle

\begin{abstract}
  The fusion of multi-source data is essential for a comprehensive analysis of geographic applications. Due to distinct data structures, the fusion process tends to encounter technical difficulties in terms of preservation of the intactness of each source data. Furthermore, a lack of generalized methods is a problem when the method is expected to be applicable in multiple resolutions, sizes, or scales of raster and vector data, to what is being processed. In this study, we propose a general algorithm of assigning features from raster data (concentrations of air pollutants) to vector components (roads represented by edges) in city maps through the iterative construction of virtual layers to expand geolocation from a city centre to boundaries in a 2D projected map. The construction follows the rule of perfect squares with a slight difference depending on the oddness or evenness of the ratio of city size to raster resolution. We demonstrate the algorithm by applying it to assign accurate PM$_{2.5}$ and NO$_{2}$ concentrations to roads in 1692 cities globally for a potential graph-based pollution analysis. This method could pave the way for agile studies on urgent climate issues by providing a generic and efficient method to accurately fuse multiple datasets of varying scales and compositions.
\end{abstract}

\begin{keywords}
  general algorithm, perfect squares, raster vector data fusion, air pollution, remote sensing, city maps
\end{keywords}

\begin{AMS}
  11Y55 11Y16 11Z05
\end{AMS}
\section{Introduction}
The fusion of multi-source remote sensing data can be classified at three different levels, pixel-level, feature-level or decision-level\cite{Zh10}. Feature-level fusion combines extracted features from distinct data sources for an integrated view. Due to the heterogeneous nature of these sources, approach towards combination are normally constrained by the characteristics of the original and target data. The most regularly used forms of data in geographic information systems are $raster$ $data$ and $vector$ $data$\cite{Huang20}. Raster data are grid-parameterized at a defined resolution and often used to represent concentrations of air pollutants. Vector data, such as graphs generated from $OpenStreetMap(OSM)$\cite{OpenStreetMap}, are composed of nodes and edges with attributes to represent road networks. A fusion process to integrate $rasters$ into $vectors$ requires assigning grid-values to graph components, nodes or edges, in vector maps as a new feature. The task can be labour-intensive and time-consuming since the source data is usually different in resolutions or sizes, and the required calculation increases greatly with the increased scale of analysis. Therefore, a general algorithm is proposed to make the fusion of these two datasets at any resolution or scale possible.

In order to read $PM_{2.5}$ values on each road in 1692 cities by checking $OSM$, the fusion between raster feature and vector maps first needs to be established, that is, to assign $PM_{2.5}$ to roads represented as edges in $OSM$ in an efficient and accurate way. Two problems will be encountered: 1) how to rasterize geographic city areas (vector maps) to the grids at the same resolution as raster pollution data and assign as expected and 2) how to make the method work with any resolution of pollution data in any sized city or number of cities.

The algorithm was developed first to find a general way of splitting the city maps into grids through perfect squares, integer sequence $A000290$\cite{A000290}, and subsequently choose an appropriate method to assign data to ensure the resulting maps preserve all original features on nodes and edges in the combined new map. These two parts are illustrated as \cref{thm:rast} and \cref{thm:findvalue} respectively in section \ref{sec:main}. The generalization is explained in \cref{thm:genera;} in the same section.



\section{Main results}
\label{sec:main}
\begin{figure}[htbp]
  \centering
  \includegraphics[width=\textwidth,height=\textheight,keepaspectratio]{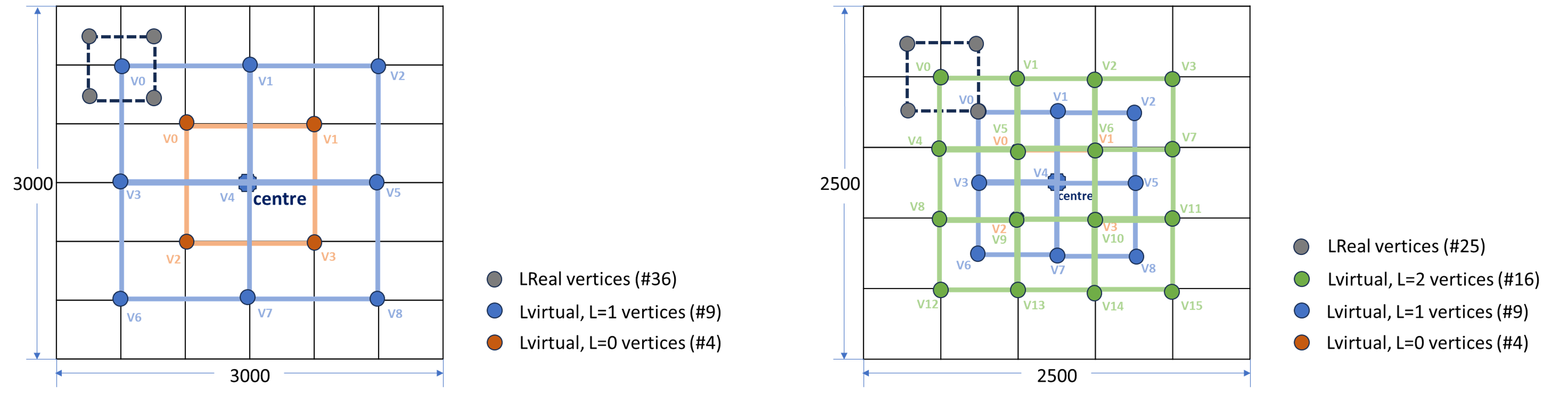}
  \caption{Map Rasterisation using Perfect Squares: step is even(left) or odd(right) number}
  \label{fig:fig1}
\end{figure}

We assume the city maps are all projected 2D squares with different sizes to be divided into smaller grids equivalent to the raster data in terms of resolution. Nodes and edges in vector maps are all indicated as geographic locations, with longitude and latitude. 

The algorithm proceeds into two sub-algorithms based on the odd or even steps it takes, moving from side to side. As shown in \cref{fig:fig1}, the number of steps $s$ are calculated by dividing city size $A$ by the target resolution $r$. For example, if resolution $r$ is 500 and city size $A$ is 3000, the number of steps $s$ is 6, which leads to even-number sub-algorithm where less iterations are required to grow spatially from the centre to the boundaries through virtually constructed layers, compared with the odd-number sub-algorithm where $A$ is 2500 and then $s$ is 5, instead. 
\begin{equation}\label{eq:step}
\DeclarePairedDelimiter{\nint}\lfloor\rceil
s=\nint[\bigg]{\frac{A}{r}}
\end{equation}
The number of  target grids $G$ is 
\begin{equation}\label{eq:grids}
G=s^{2}
\end{equation}
For even-numbered sub-algorithm, the number of total iterations $I_{totale}$ is
\begin{equation}\label{eq:iter-even}
I_{totale}=\sqrt{\frac{G}{4}}=\frac{s}{2}
\end{equation}
For odd-numbered sub-algorithm, $I_{totalo}$ is
\begin{equation}\label{eq:iter-odd}
I_{totalo}=\sqrt{G}-1=s-1
\end{equation}
The number of total iterations is represented as total layers in \cref{fig:fig1}, where $I_{totale}$ is 3 and $I_{totalo}$ is 4. The last layer in gray is defined as real since it produces the final list of centers of target grids. Layers above the last are named as virtual since they are used to expand spatially in a gradually repetitive way to arrive at the last layer. So, the number of virtual iterations $I_{vire(o)}$ is 
\begin{equation}\label{eq:iter-virtual}
I_{vire(o)}=I_{totale(o)}-1
\end{equation}
The single-side sizes of real grids in the last layer is,
\DeclarePairedDelimiter{\nint}\lfloor\rceil
\begin{equation}\label{eq:brs}
B_{reale(o)}=\nint[\bigg]{\frac{A}{s}}
\end{equation}
And the single-side sizes of virtual grids (colored squares in \cref{fig:fig1}) $B_{vire}$ is 
\begin{equation}\label{eq:bs-even}
B_{vire}=2\times B_{reale}
\end{equation}
for $s$ is even; $B_{viro}$ is
\begin{equation}\label{eq:bs-odd}
B_{viro}=B_{realo}
\end{equation}
for $s$ is odd. 

\begin{theorem}[Map Rasterisation using Perfect Squares]\label{thm:rast}
If $I_{totale(o)} > 1$, the number of distinct vertices $V_{vire(o)}$ of constructed virtual grids with the areas of $B_{vire(o)}^2$  expands through $I_{vire(o)}$ iterations as perfect squares started from 4. The number of distinct vertices $V_{reale(o)}$ of real grids with the areas of $B_{reale(o)}^2$ differ in the calculations of even or odd steps $s$. If $I_{totale(o)} = 1$, no virtual grid is required and $V_{reale(o)}$ is 4. If $I_{totale(o)} = 0$, no rasterisation is required.\newline
If $I_{totale(o)} > 1$:
\begin{displaymath}
  V_{vire(o)} = \{4,9,16,\ldots ,(n+2)^{2}\}, n = 0,1,2,\dots ,I_{vire(o)}-1 
  \end{displaymath}
  for s is even:
  \begin{displaymath}  
    V_{reale} = 4\times (n+2)^{2}
  \end{displaymath}  
  for s is odd:
  \begin{displaymath}  
    V_{realo} = ((n+1)+2)^{2}   
  \end{displaymath}
  $V_{reale(o)}$ should be equal to the number of target grids $G$, 
  \begin{displaymath}  
    V_{reale(o)} == G   
\end{displaymath}
\end{theorem}

The iterations proceed by converting centres to vertices repeatedly, and then the vertices are used as new centres to generate new vertices without duplicates, continuing until the centres of the target grids are all obtained\textemdash resulting in a list of geolocations indicated as latitudes and longitudes $G(lat,lon)$. Next, raster data $D_r$, as features as concentrations of air pollutants $P$, is to be assigned into vector maps $D_v$ based on these geolocations. The mapping of two sets of geolocations is processed through bounding boxes with the areas of $B_{reale(o)}^2$. 


\begin{theorem}[Assign Raster Features to Vector Maps]\label{thm:findvalue}
  Suppose raster data $D_{r}$, including the feature of interest $P$, is continuous 
  and present everywhere, as in vector map data $D_{v}$, 
  the geographic locations $G(lat,lon)$ of the created centres in \cref{thm:rast} are used to retrieve pollution data $P$ in $D_{r}$ at resolution $B_{reale(o)}$ and is converted as $B_{geo}$ to align units with $G(lat,lon)$. Then, $P$ extends $D_{v}$.
  \begin{displaymath}
    P(lat,lon)=D_{r}(P(lat^\prime,lon^\prime)) 
  \end{displaymath}
  in which,
  \begin{displaymath}
    lat^\prime\in[lat-\frac{1}{2}B_{geo},lat+\frac{1}{2}B_{geo}], lon^\prime\in[lon-\frac{1}{2}B_{geo},lon+\frac{1}{2}B_{geo}].
  \end{displaymath}
  Vector map data is updated as,
  \begin{displaymath}
    D_{v}^{(1)}=D_{v}^{(0)}\cup P
  \end{displaymath}
\end{theorem}

The algorithm can be easily generalized on two levels: 1) raster data can be at any resolution, accommodating observations at any scale, and vector maps can be any size for varied urban morphologies of cities, globally. 2) large-scale analysis can be achieved by increased iterations. 
\begin{theorem}[General Algorithm at Any Resolution or Scale]\label{thm:genera;}
  Apply \cref{thm:rast} and \cref{thm:findvalue} to any resolution of raster data by changing $r$, any size of vector maps by changing $A$ in \cref{eq:step} followed with updated \cref{eq:grids,eq:iter-even,eq:iter-odd,eq:iter-virtual,eq:bs-even,eq:bs-odd,eq:brs}, and to any number of cities via required number of iterations.
\end{theorem}





\section{Algorithm}
\label{sec:alg}
The algorithm  for single city is stated in \cref{alg:31} and \cref{alg:32} to realize \cref{thm:rast} and \cref{thm:findvalue} respectively. The generalization of \cref{thm:genera;} for any number of cities is conducted through \cref{alg:33}. Raw codes can be found in \cref{fig:fig7} and \cref{fig:fig8} in the supplementary material.

\begin{algorithm}
\caption{for \cref{thm:findvalue}}
\label{alg:32}
\begin{algorithmic}
\STATE{Define $data_{vector}:=D_v, data_{raster}:=D_r, value_{feature}:=P,attributes_{edges}:=E_{attr},nodes_{global}:=N_{glb},edges_{global}:=E_{glb},nodes_{local}:=N_{loc},edges_{local}:=E_{loc}$}
\STATE{$N_{glb},E_{glb}\Leftarrow D_v^{(0)}(center_{vector},A)$}
\STATE{$E_{sum}=[$ $]$}
\FOR{$pos \in latlon$}
\STATE{$N_{loc},E_{loc}\Leftarrow D_v(pos,B_{reale(o)})$}
\STATE{$P:=D_r(pos,B_{geo})$}
\STATE{$E_{attr}\mid_{E_{loc}}\gets P$}
\STATE{Update $E_{loc}$}
\STATE{$E_{sum}:=E_{sum}\cup E_{loc}$}
\ENDFOR
\STATE{Check if $\# E_{glb} == \# E_{sum}$}
\STATE{$D_v^{(1)}\Leftarrow N_{glb},E_{sum}$}
\RETURN $D_v^{(1)}$
\end{algorithmic}
\end{algorithm}

\begin{algorithm}
\caption{for \cref{thm:genera;}}
\label{alg:33}
\begin{algorithmic}
\STATE{Define $list_{city}:=C, D_r$}
\FOR{$city \in C$}
\STATE{$center_{vector},A\Leftarrow city$}
\STATE{$r\Leftarrow D_r$}
\STATE{Run \cref{alg:31}}
\STATE{Run \cref{alg:32}}
\ENDFOR
\end{algorithmic}
\end{algorithm}
\begin{algorithm}
\caption{for \cref{thm:rast}}
\label{alg:31}
\begin{algorithmic}
\STATE{Define $center_{vector}:=(lat,lon), size_{vector}:=A, resolution_{raster}:=r$}
\STATE{$s := \lfloor{\frac{A}{r}}\rceil$}
\STATE{$G := s^2 $}
\STATE{$latlon := [(lat,lon)]$}
\WHILE{$\#s > 1$}
\IF{$s$ is even}
 \STATE{$I_{totale}:=\sqrt{\frac{G}{4}}:=\frac{s}{2}$}
 \STATE{$list_{latlon}:=[(lat,lon)]$}
 \WHILE{$\#I_{totale}>1$} 
 \STATE{$I_{vire}:=I_{totale}-1,B_{reale}:=\lfloor{\frac{A}{s}}\rceil,B_{vire}:=2\times B_{reale}$}
 \STATE{Define $list_{latlon}=$$[$ $]$} \STATE{$[north_0^{(0)},east_0^{(0)},west_0^{(0)},south_0^{(0)}]\Leftarrow center_{vector},B_{vire}$}
 \STATE{$list_{latlon}:=list_{latlon}\cup(north_0^{(0)},east_0^{(0)})
 \cup(north_0^{(0)},west_0^{(0)})$\\$\cup(south_0^{(0)},east_0^{(0)})\cup(south_0^{(0)},west_0^{(0)})$} 
 \FOR{$i \gets 0$ to $I_{vire}$}
 \STATE{Define $list_{latlon}=$$[$ $]$}
 \FOR{$k \gets 0$ to $4^{i}$}
\STATE{$[north_{4k}^{(i+1)},east_{4k}^{(i+1)},west_{4k}^{(i+1)},south_{4k}^{(i+1)}]\Leftarrow (north_k^{(i)},east_k^{(i)}),B_{vire}$}
\STATE{$[north_{4k+1}^{(i+1)},east_{4k+1}^{(i+1)},west_{4k+1}^{(i+1)},south_{4k+1}^{(i+1)}]\Leftarrow (north_k^{(i)},west_k^{(i)}),B_{vire}$}
\STATE{$[north_{4k+2}^{(i+1)},east_{4k+2}^{(i+1)},west_{4k+2}^{(i+1)},south_{4k+2}^{(i+1)}]\Leftarrow (south_k^{(i)},east_k^{(i)}),B_{vire}$}
\STATE{$[north_{4k+3}^{(i+1)},east_{4k+3}^{(i+1)},west_{4k+3}^{(i+1)},south_{4k+3}^{(i+1)}]\Leftarrow (south_k^{(i)},west_k^{(i)}),B_{vire}$}
 \ENDFOR
 \STATE{$list_{latlon}:=list_{latlon}\cup (north_{0,1,\dots,4^{i+1}-1}^{(i+1)},east_{0,1,\dots,4^{i+1}-1}^{(i+1)})$} 
 \STATE{$list_{latlon}:=list_{latlon}\cup (north_{0,1,\dots,4^{i+1}-1}^{(i+1)},west_{0,1,\dots,4^{i+1}-1}^{(i+1)})$} 
 \STATE{$list_{latlon}:=list_{latlon}\cup (south_{0,1,\dots,4^{i+1}-1}^{(i+1)},east_{0,1,\dots,4^{i+1}-1}^{(i+1)})$} 
 \STATE{$list_{latlon}:=list_{latlon}\cup (south_{0,1,\dots,4^{i+1}-1}^{(i+1)},west_{0,1,\dots,4^{i+1}-1}^{(i+1)})$} 
 \ENDFOR
 \STATE{Remove duplicates in $list_{latlon}$}
 \ENDWHILE
\ELSIF{$s$ is odd}
 \STATE{$I_{totalo}:=\sqrt{G}-1:=s-1$}
 \WHILE{$\#I_{totalo}>1$} \STATE{$I_{viro}:=I_{totalo}-1,B_{realo}:=\lfloor{\frac{A}{s}}\rceil,B_{viro}:=B_{realo}$}
 \STATE{Apply the same part of the codes as in $even$ by replacing $I_{vire},B_{vire},B_{reale}$ with $I_{viro},B_{viro},B_{realo}$}
 \ENDWHILE
\ENDIF
\STATE{Define $latlon=$$[$ $]$}
 \FOR{ $item \in list_{latlon}$}
 \STATE{$[north,east,west,south]\Leftarrow item,B_{reale(o)}$}
 \STATE{$latlon:=latlon\cup(north,east)
 \cup(north,west)$\\$\cup(south,east)\cup(south,west)$}
 \ENDFOR
 \STATE{Remove duplicates in $latlon$}
\STATE{Check if $\#latlon == G$}
\ENDWHILE
\RETURN $latlon$
\end{algorithmic}
\end{algorithm}


\section{Experimental results}
\label{sec:experiments}
\begin{figure}[htbp]
  \centering \includegraphics[width=\textwidth,height=\textheight,keepaspectratio]{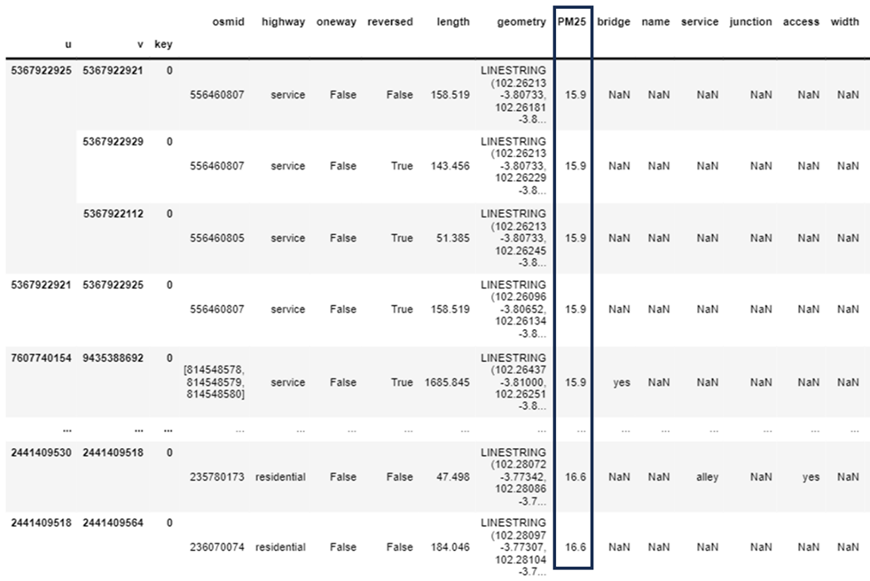}
  \caption{$PM_{2.5}$ added into $OSM$}
  \label{fig:fig5}
\end{figure}

\begin{figure}[htbp]
  \centering \includegraphics[width=\textwidth,height=\textheight,keepaspectratio]{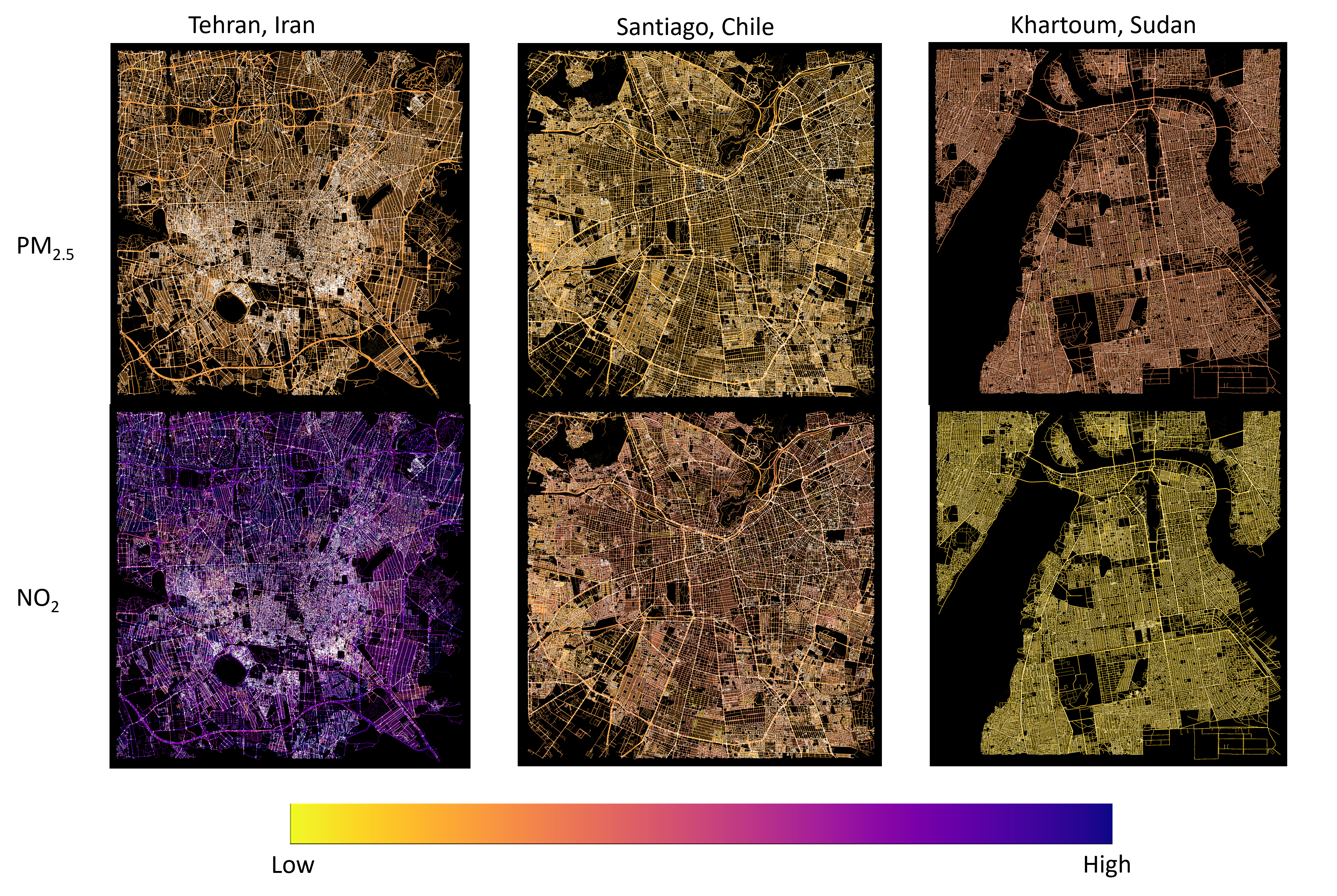}
  \caption{Sample images of $PM_{2.5}$ and $NO_{2}$ shown on each road in 1692 cities}
  \label{fig:fig6}
\end{figure}

We applied the stated algorithms to the fusion of two raster datasets of $PM_{2.5}$\cite{pm25} and $NO_{2}$\cite{no2}, and vector map data generated from $OSM$\cite{OpenStreetMap} for 1692\footnote{$NO_{2}$ is only available for 1689 cities resulted from its source data.} global cities. The city list including information of centers and sizes has been used in the previous study from our group\cite{citylist}. \cref{fig:fig5} shows $PM_{2.5}$ as a new attribute being added into edges columns of $OSM$\cite{OpenStreetMap} through intermediate step \cref{fig:fig3}. As a result,$PM_{2.5}$ and $NO_{2}$ are on each road in 1692 cities as sampled in \cref{fig:fig6}.





\section{Discussion}
All discussed centers and vertices are indicated as decimals, and duplicates are identified on the same appropriate decimal places. In this study, 3 decimal places work for most cities, but 4 is required for $\sim1\%$ cases in the 1692 cities. Trial number of places was set to avoid failure. Alternatively, trail iterations to find that $\sim1\%$ run first and a new number is then set for specific items. Correctness checks on whether $G==\#latlon$ are conducted after every iteration of \cref{alg:31} with output as in \cref{fig:fig2}. The algorithm could be improved by encoding decimals with integers for duplicates-check followed by the same-reverse-path decoding. 

Furthermore, we initially tried to merge the $E_{loc}$ to $E_{sum}$ based on graphs with the process visualized in \cref{fig:fig4}. Edges that cross boundaries are cut into segments. The method is however found to produce breaks on edges in the merged graphs and unsuitable as the breaks changed the topology of the original graphs. Therefore, we treated edges as frame-data to rasterise and remerge, as in \cref{alg:32}, instead. All original properties on graphs and attributes from nodes and edges are preserved. 

Finally, we assumed all cities are represented as square vector maps in our study, but they are normally not square in reality. This algorithm should then be followed by additional steps to process variant shapes of boundaries.

\section{Conclusions}
\label{sec:conclusions}
The feature-level fusion of raster and vector data can be efficiently achieved at any resolution, size or scale by utilizing the proposed algorithms, by which the technique was represented  has application when integrating multiple sources of data at large scale is required. In this example, we illustrate the technique using data related to air pollution across cities, globally, thereby providing valuable insights on climate change and its dynamic trajectory. The technique is low-cost and convenient to adapt to other forms of fusions on pixel or decision levels.



\section*{Acknowledgments}
The first author is supported by Melbourne Research Scholarship.

\bibliographystyle{siamplain}
\bibliography{references}
\end{document}


\maketitle

\section{Raw codes}

\begin{figure}[htbp]
  \centering
  \includegraphics[width=\textwidth,height=\textheight,keepaspectratio]{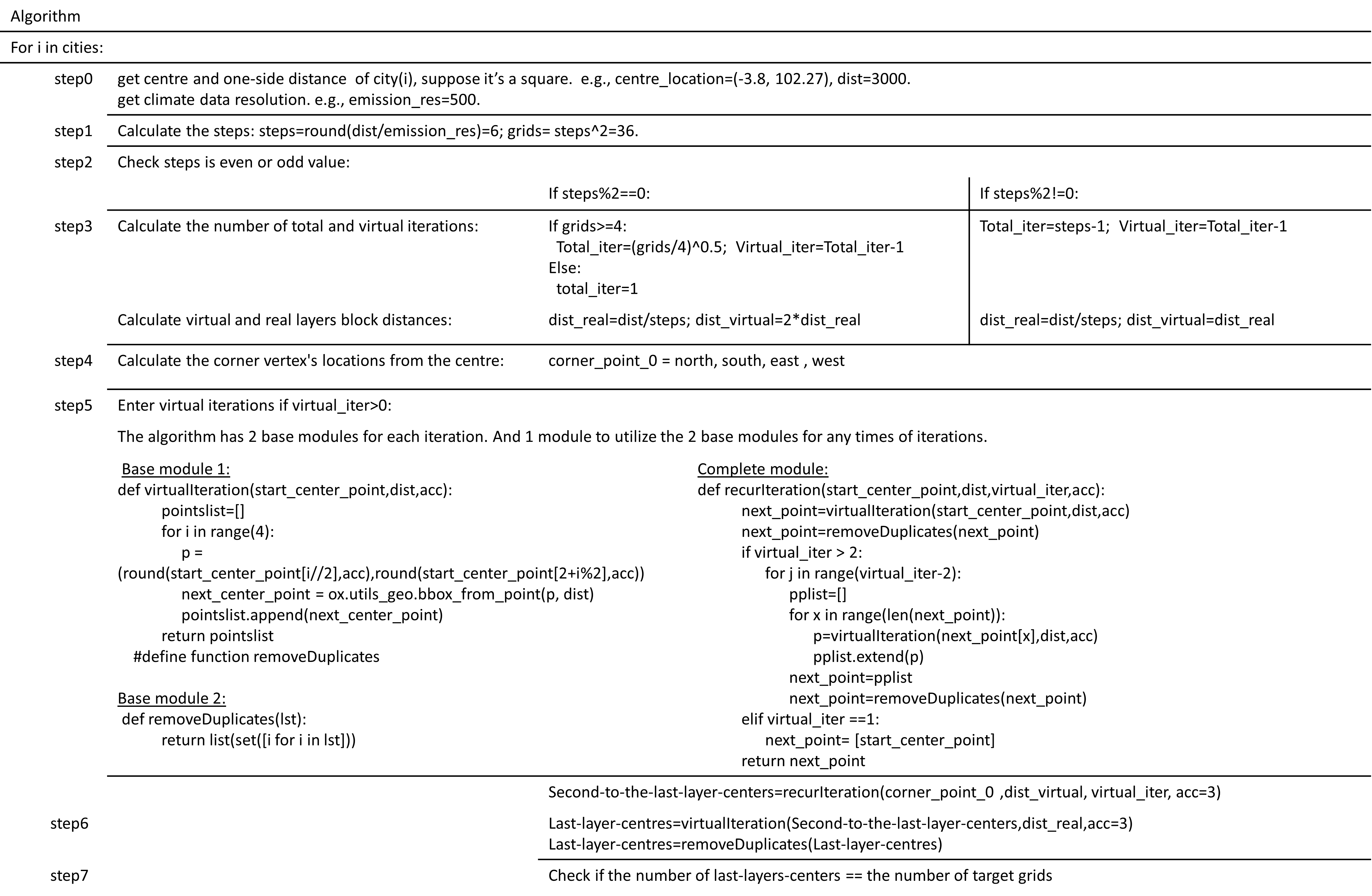}
  \caption{Codes for Map Rasterisation described in \cref{thm:rast}}
  \label{fig:fig7}
\end{figure}

\begin{figure}[htbp]
  \centering
  \includegraphics[width=\textwidth,height=\textheight,keepaspectratio]{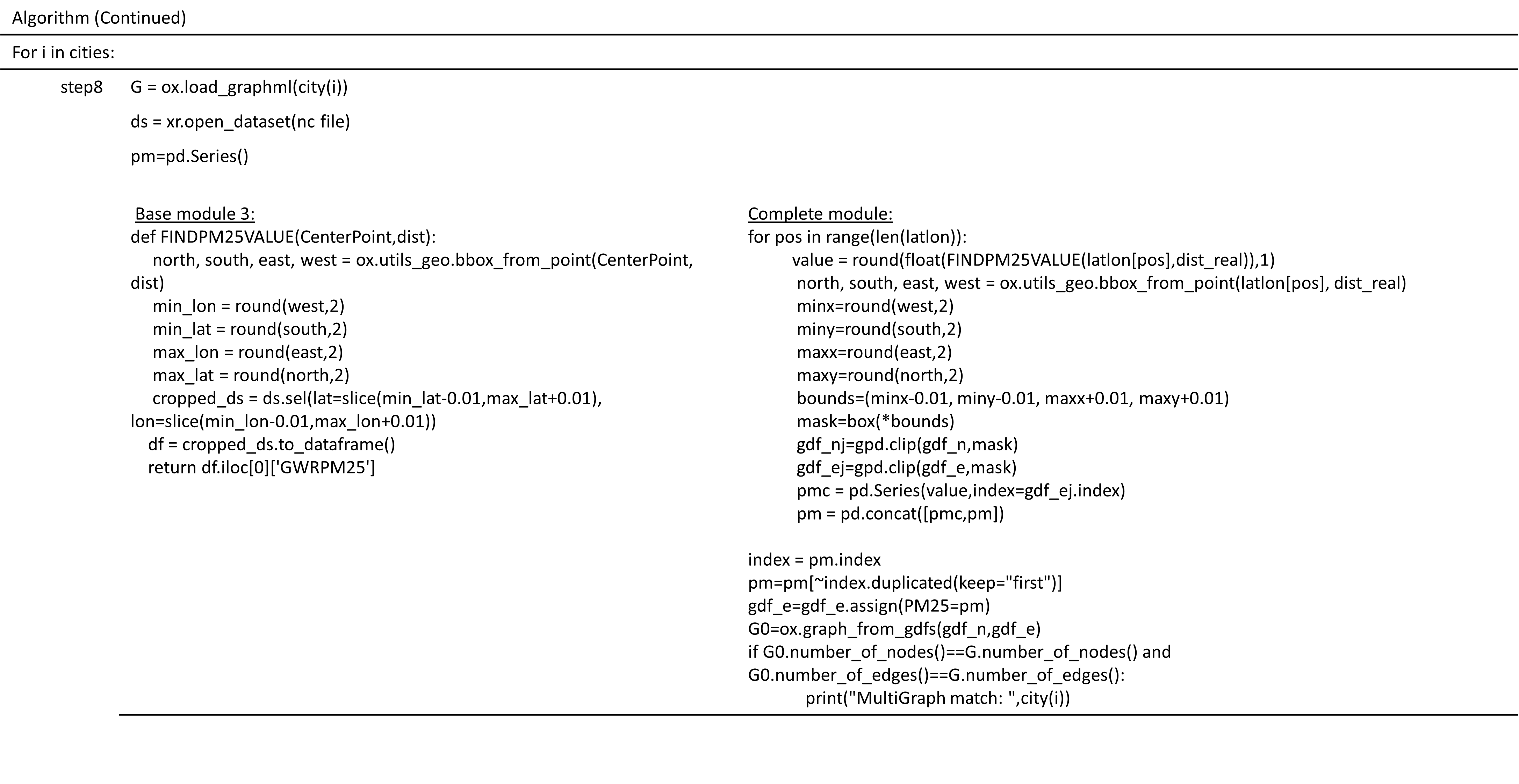}
  \caption{Codes for Assigning Data described in \cref{thm:findvalue}}
  \label{fig:fig8}
\end{figure}

Here we include the raw codes for \cref{alg:31} and \cref{alg:32} in \cref{fig:fig7} and \cref{fig:fig8} respectively for your reference. Variable names are slightly different on both sides.

\section{Supplementary figures}
\begin{figure}[htbp]
  \centering  \includegraphics[width=\textwidth,height=\textheight,keepaspectratio]{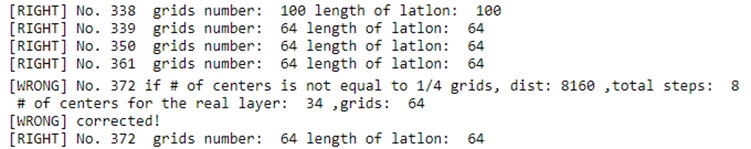}
  \caption{Sample of output with correction of accuracy setting}
  \label{fig:fig2}
\end{figure}

Sample of output in \cref{fig:fig2} is the result of checking whether $G==\#latlon$ after \cref{alg:31}. Error is corrected by modifying the number of decimal places from 3 to 4.

\begin{figure}[htbp]
  \centering
\includegraphics[width=\textwidth,height=\textheight,keepaspectratio]{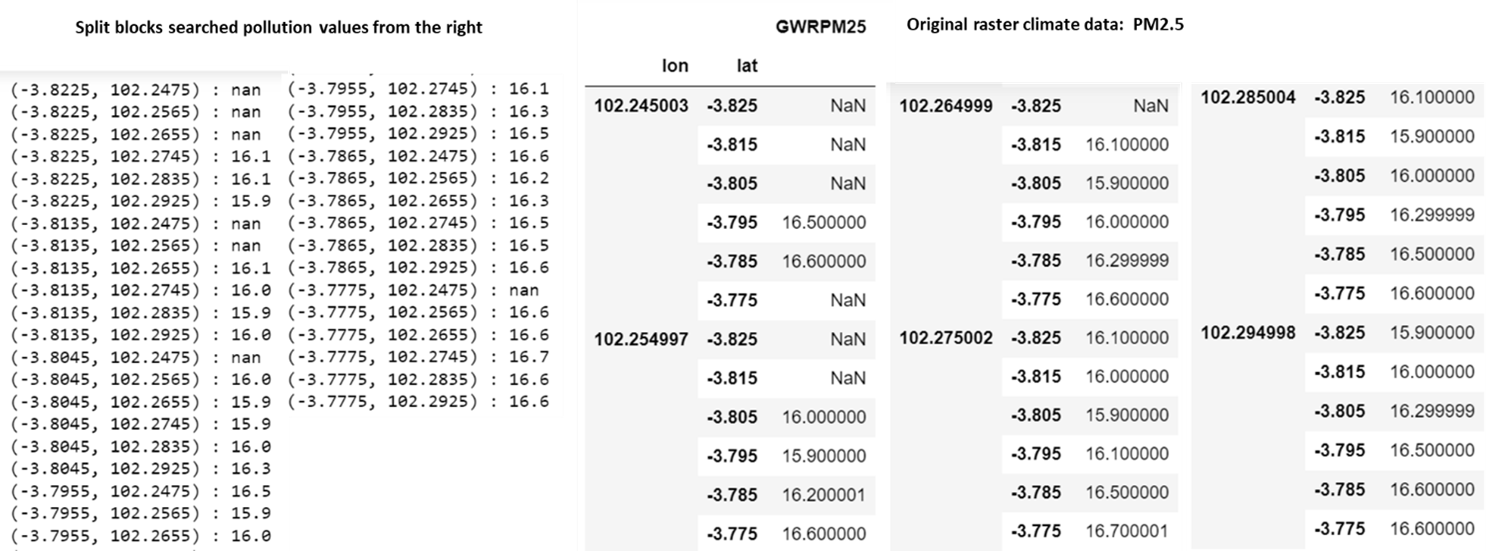}
 \caption{Sample of output of \cref{thm:rast} and \cref{thm:findvalue}}
  \label{fig:fig3}
\end{figure}

\cref{fig:fig3} is a sample output of $map$ $rasterisation$ into $36$ grids with centres listed left in the $(lat, lon)$ format. $PM2.5$ values are queried from the right original raster data.

\begin{figure}[htbp]
  \centering
  \includegraphics[width=\textwidth,height=\textheight,keepaspectratio]{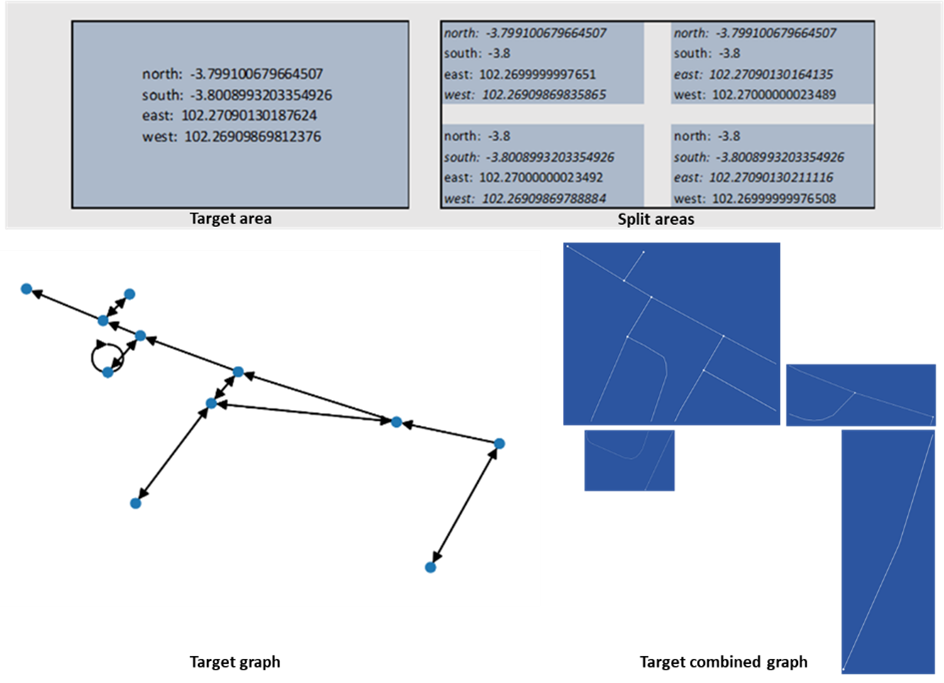}
  \caption{Visualization of vector components rasterisation and remerging}
  \label{fig:fig4}
\end{figure}

The graphs in \cref{fig:fig4} visualize the process of \cref{alg:32} if graph-approach is used. The original graph ($bottom-left$) is rasterised into $4$ smaller graphs($bottom-right$) according to the boundary boxes($top$).


%% file: ex_shared.tex
\usepackage{lipsum}
\usepackage{amsfonts}
\usepackage{graphicx}
\usepackage{epstopdf}
\usepackage{algorithmic}
\ifpdf
  \DeclareGraphicsExtensions{.eps,.pdf,.png,.jpg}
\else
  \DeclareGraphicsExtensions{.eps}
\fi

\newsiamremark{remark}{Remark}
\newsiamremark{hypothesis}{Hypothesis}
\crefname{hypothesis}{Hypothesis}{Hypotheses}
\newsiamthm{claim}{Claim}

\headers{General alg. of assig. r. feat. to v. maps at any res. or scale}{N. Xu, M. Stevenson, K. Nice and S. Seneviratne}

\title{General algorithm of assigning raster features to vector maps at any resolution or scale\thanks{Submitted to the editors May 25, 2024.
}}

\author{Nan Xu\thanks{Department of Infrastructure Engineering, Faculty of Engineering and IT, University of Melbourne, Victoria, Australia
  \email{naxu@student.unimelb.edu.au}
  }
\and 
Mark Stevenson\footnotemark[2]
\footnotemark[4]
\thanks{Centre for Epidemiology and Biostatistics, Melbourne School of Population and Global Health, University of Melbourne, Victoria, Australia
\email{mark.stevenson@unimelb.edu.au}
}
\and 
{Kerry A. Nice}\thanks{Transport, Health and Urban Systems Research Lab, Faculty of Architecture, Building and Planning, University of Melbourne, Victoria, Australia \email{kerry.nice@unimelb.edu.au} }
\and 
{Sachith Seneviratne}\footnotemark[4]
\thanks{Department of Mechanical Engineering, Faculty of Engineering and IT, University of Melbourne, Victoria, Australia
\email{sachith.seneviratne@unimelb.edu.au}
}
}

\usepackage{amsopn}

\makeatletter
\newcommand*{\addFileDependency}[1]{
  \typeout{(#1)}
  \@addtofilelist{#1}
  \IfFileExists{#1}{}{\typeout{No file #1.}}
}
\makeatother

\newcommand*{\myexternaldocument}[1]{%
    \externaldocument{#1}%
    \addFileDependency{#1.tex}%
    \addFileDependency{#1.aux}%
}